\documentstyle[preprint,prb,aps]{revtex}
\begin{document}
\draft

\title {Measurement of the conductance distribution function at 
a quantum Hall transition}

\author{D. H. Cobden$^{\dagger}$\cite{Berkeley} and E. Kogan$^{\S}$}

\address{$^{\dagger}$Cavendish Laboratory, Madingley Road, Cambridge
CB3 0HE, United Kingdom\\
$^{\S}$Jack and Pearl Resnick Institute of Advanced Technology,\\
Department of Physics, Bar-Ilan University, Ramat-Gan 52900, 
Israel}

\date{\today }

\maketitle

\begin{abstract}
We study experimentally the reproducible conductance fluctuations between the
quantum Hall plateaus in the conductance of two-terminal submicron silicon
MOSFETs.  For the dramatic fluctuations at the insulator-to-first-plateau
transition we find a conductance distribution that is approximately uniform
between zero and $e^2/h$.  We point out that this is consistent with the 
prediction of random $S$-matrix theory for a conductor with single-channel 
leads in a magnetic field. 
\end{abstract}

\pacs{PACS numbers: 73.23.-b, 73.23.Ad, 73.40.Hm}

Reproducible fluctuations in the conductance of mesoscopic electronic
devices have been a subject of continued interest for more than a decade
\cite{bee}.  At low magnetic fields they are now rather well understood in
the case of diffusive conductors \cite{alt}, while their manifestation in
chaotic ballistic cavities and quantum dots is under intensive study
\cite{mel,bro} at present.  However, relatively little is known about the
fluctuations that occur in high magnetic fields.  For a two-dimensional
system in which the quantum Hall effect dominates, the fluctuations occur
whenever the conductivity is not quantized \cite{tim,for,jai,sim,mai},
i.e. in the transition regions, where the physics is tremendously subtle
\cite{pra}. 

At present, two complementary microscopic pictures exist of the cause of
the fluctuations at high field.  In one picture, transport near the center
of the transition is assumed to be in the diffusive limit, with
modifications to the diffusion equation to allow for Landau quantization
and boundary effects \cite{sha,mas,khm}.  In this case the fluctuations
resemble those at zero magnetic field $B$, but with correlation scales
that vary with $B$, as has been observed in experiments \cite{gei,mor}. 
This picture does not take into acount localization (and therefore the
quantum Hall effect), and is applicable only when the index of the highest
occupied Landau level is large.  In the other picture, transport is in the
adiabatic limit, and peaks in the conductance or resistance arise from
resonant tunneling between edge states through localized bulk states
\cite{jai}.  This picture is applicable if the disorder potential is
smooth on the scale of the magnetic length, so that edge states are well
defined within the sample.  It has been used to interpret fluctuations at
the edges of quantum Hall zeros in the four-terminal longitudinal
resistance of high mobility GaAs/AlGaAs heterostructure devices
\cite{sim,mai}.  However, one is faced with the problem that in the
archetypal integer quantum Hall system, the silicon MOSFET
(metal-oxide-silicon field-effect transistor), which we study here, the
conditions are such that neither of these two limiting cases pertains.  In
MOSFETs the disorder is very short range, while wide integer quantum Hall
plateaus are seen at low Landau-level index.  This, combined with the
unknown but certainly important consequences of interactions, makes it
very difficult to interpret the dramatic fluctuations we observe in our
devices in terms of a microscopic model at present. 

Nonetheless, the techniques of random matrix theory (RMT) allow one to
make predictions for the statistical properties of conductance
fluctuations without knowing all the microscopic details.  RMT has
recently been successfully applied to diffusive transport in small
metallic objects and ballistic transport in chaotic cavities \cite{sto}. 
Here we make the suggestion that under certain conditions RMT may also be
applied to transport at quantum Hall transitions, as described by the
scattering matrix between incoming and outgoing edge channels \cite{bee}. 
We present experimental evidence in support of this, namely that the
measured distribution of the fluctuations on a quantum Hall transition is
approximately flat between zero and $e^2/h$.  The same flat distribution
is predicted by RMT for the single-channel $S$-matrix at high $B$
\cite{har,jal}.  In contrast, at low $B$ we observe the Gaussian
distribution expected when the conductance (and therefore the number of
channels) is large \cite{alt}. 

The devices used are industrial process two-terminal silicon MOSFETs with
oxide thickness $250$~\AA‰ and a range of effective channel dimensions
down to $0.4$~$\mu$m.  The rough sketch in the inset to Figure 1
emphasizes their essentially very simple geometry.  The n+ contacts,
indicated in the plan view by rectangles with diagonal crosses, are
self-aligned to the gate.  Each device can be thought of as a rectangle of
two-dimensional electron gas with an almost ideal metallic contact at each
end and a variable carrier density controlled linearly by the gate voltage
$V_g$.  The conductance was measured in a dilution refrigerator using a
low-frequency ac voltage of $10$~$\mu$V (comparable with the effective
electron temperature of approximately $100$~mK), a virtual-earth current
preamplifier and a lock-in amplifier. The low temperature mobility was
around $2000$~cm$^2$~V$^{-1}$~s$^{-1}$, corresponding to a mean free path
of about $200$~\AA.  At base temperature the phase coherence length was
around $0.4$~$\mu$m, as deduced from the low-$B$ fluctuations (see below)
\cite{sup}. 

The characteristics of a $0.6 \times 0.6$~$\mu$m$^2$ device are shown in
Figure 1(a) at $B = 0$ and Figure 1(b) at $B = 16$~T.  At $B=0$ the
conductance $G$ as a function of $V_g$ shows universal fluctuations
superimposed on a rising background above a threshold of about $2.2$~V
(this threshold varies randomly between devices over a range of $500$~mV
or so).  At $16$~T the situation is very different: the conductance
increases through a series of quantum Hall plateaus.  This is a
consequence of the fact that for such a nearly ideal two-terminal device
the Landauer formula (Eq. (2) below) applies directly to the measured
conductance.  On the plateaus, the conductance (in units of $e^2/h$)
equals the number of transmitted edge channels.  (Note that previous
studies of fluctuations at high magnetic fields
\cite{tim,for,sim,mai,gei,mor} employed multi-terminal Hall-bar geometries
for which the relation of the measured quantities to the transmission
coefficients is less direct).  Subtraction of a constant series resistance
was found to bring all the plateaus very close to exact multiples of
$e^2/h$.  Because of the double spin and valley degeneracies at zero
field, the first four plateaus are all associated with the lowest Landau
level.  On the transitions between adjacent plateaus there are rapid,
reproducible fluctuations. 

Before discussing these fluctuations at high $B$ further, let us compare
those at low $B$ with the existing standard theory \cite{alt}.  Below
about $4$~T, sweeping $B$ generates fluctuations, as shown in Figure 2(a),
which are statistically independent of those produced by sweeping $V_g$
(as in Figure 1(a)).  The magnetic field correlation length \cite{alt2} is
$B_c \approx 30$~mT, implying that the phase-coherent area is $h/(eB_c)
\approx 0.14$~$\mu$m$^2$, which is more than a third of the channel area,
$0.36$~$\mu$m$^2$.  To obtain the conductance distribution function,
similar $B$-sweeps were taken at a series of closely spaced values of
$V_g$ between $3.6$ and $4.0$~V and a smooth monotonic background was
subtracted from the $G$-$B$-$V_g$ dataset.  The histogram of the resulting
conductance values, offset by the average, $\bar G = 17.5$~$e^2 /h$, of
$G$ over the entire dataset, is shown in Figure 2(b).  Since $\bar G \gg
1$, a purely Gaussian distribution is expected \cite{alt} for a
phase-coherent sample.  The solid curve in the figure is the best fit
Gaussian.  Of course, for a highly phase-incoherent sample, effectively
consisting of a large number of uncorrelated phase-coherent units, the
central limit theorem will enforce a Gaussian distribution.  However,
there are fewer than three phase-coherent sub-units in this device, and we
believe that the accurate measured Gaussianity reflects the intrinsic
conductance distribution.  The standard deviation of the measured
distribution is $0.22$~$e^2/h$, which taking into account partial
phase-breaking, uncertain geometrical factors, and an unknown contact
resistance, is quite consistent with the measured value of $B_c$,
according to the theory of universal conductance fluctuations
\cite{bee,alt}. 

We now turn to the high magnetic field results.  Figure 2(c) shows the
lowest quantum Hall transition from Figure 1(b).  The thicker solid line
superimposed is the same data smoothed by averaging over $16$~mV in $V_g$. 
At the edges of the transition the fluctuations in $G$ resemble the
resonant peaks and dips of the edge-state tunneling picture
\cite{jai,sim,mai}.  However, over a wide region of $V_g$, indicated by
plotting the data using a thin solid line rather than a dashed line, $G$
fluctuates strongly between strict limits of zero and $e^2/h$. Throughout
this region the smoothed conductance lies in the range $0.5 \pm
0.1$~$e^2/h$.  Figure 2(d) shows the distribution of conductance values
for this region only.  It is fairly uniform for $0 < G < e^2/h$ and zero
for $G > e^2/h$.  This contrasts sharply with the strongly peaked,
long-tailed Gaussian distribution for $B=0$ in Figure 2(b). For reasons
which will be discussed below we have no way of improving the statistics
here.  However, we can estimate the error in the value for each bin in the
histogram to be of the order of the variation from bin to bin.  To within
this error the distribution is flat between $0.1$ and $0.9$~$e^2/h$. 

Two versions of RMT have been used in transport theory.  The first, that
of random multiplicative transfer matrices \cite{sto}, can describe
diffusive transport in quasi-one-dimensional systems.  The second, that of
the random $S$-matrix, is most appropriate for ballistic transport.  Here
we are dealing with a third kind of transport, which we imagine involves
the adiabatic motion of electrons in rather long, percolating edge-like
quantum Hall states.  Taking into account also the strong disorder in the
system, one might guess that this problem belongs in the same universality
class as diffusive transport.  However, the main aim of this paper is to
point out that a flat distribution of the conductance between zero and
$e^2/h$, compatible with what we observe for the first quantum Hall
transition, is just what one gets from the simplest random $S$-matrix
theory. 

The appropriate RMT result has already been obtained in the
analysis of chaotic ballistic cavities \cite{har,jal}.  For a sample
connected between two leads each with $N$ channels, the $S$-matrix is
\begin{equation}
\label{1}
S=\left(\matrix{ r & t' \cr
t & r'\cr}\right),
\end{equation}
where $r$ and $t$ are the $N \times N$ reflection and transmission
matrices for electrons incident from the left, and $r'$ and $t'$ are for
those from right.  Current conservation forces $S$ to be unitary, and
since the strong magnetic field breaks time reversal symmetry the unitary
ensemble, with symmetry index $\beta = 2$, is appropriate.  The
two-terminal conductance is given by the Landauer formula,
\begin{equation}
\label{2}
G=(e^2/h)Tr\{t t^{\dagger}\}.
\end{equation}

For a single quantum Hall transition, and especially the first (between
$G=0$ and $e^2/h$), it is natural to take $N=1$.  The $2 \times 2$
$S$-matrix then consists of the reflection and transmission coefficients
for the lowest (spin-down, valley-one) edge state incident on the 2D
region from the contacts.  If $S$ is characterized by Dyson's circular
ensemble then the conductance distribution function is given by
\cite{har,jal}
\begin{equation}
\label{3}
P(G) = (\beta/2)G^{-1+\beta/2}.
\end{equation}
For $\beta = 2$, Eq. (3) gives a flat distribution between zero and
$e^2/h$, with $\bar G = 0.5$~$e^2/h$.  The data in Fig. 2 (c) are
particularly appropriate for comparison with this theory because of the
wide range of $V_g$ over which the average of $G$ (the thick solid line)
is close to $0.5$~$e^2/h$. 

We emphasize that the form of the fluctuations on quantum Hall transitions
is variable, and it is difficult to find instances like the one in Figure
2(b) where the data are suitable for obtaining a distribution function. 
This is illustrated by the examples of transitions shown in Figure 3.  One
inhibiting factor is that the correlation length, or period, of the
fluctuations is usually longer than in this device.  The period, and hence
the number of peaks and dips, can in fact differ markedly between devices,
as illustrated by trace (iii) from device 2, which is nominally identical
to device 1.  A difference in period and amplitude may be explained by
variations in the electron temperature, due to changes in the amount of
heating by external electrical noise.  However, temperature variations
cannot account for the reproducible differences in period often seen
between transitions in the same device, and even across a single
transition, such as in trace (i).  The monotonic increase in $\bar G$ and
the uniformity of the fluctuations actually appear to be restricted to
square devices and to the first quantum Hall transition only.  For higher
transitions, as illustrated by traces (i) and (ii) which are again from
device 1, the potential for intervalley and spin-flip scattering may
underlie the complex nonmonotonic behavior.  Meanwhile, when the
length/width ratio differs from unity, as it does for device 3 (trace
(iv)), one can easily imagine the asymmetry between the average forward-
and backward-scattering probabilities resulting in a situation where the
$S$-matrix is not completely random \cite{mel}. 

Nevertheless, we can discern the following patterns.  First, as the
device area increases, the period in $V_g$ of the fluctuations tends to
decrease.  Second, in devices where the length/width ratio is much less
than or greater than unity, the valley-split plateaus are destroyed by
fluctuations with respectively an enhanced or a reduced mean conductance,
while the other plateaus are remarkably robust.  Third, the fluctuations
evolve in a surprisingly simple way with $B$ \cite{bev}. These results
will be reported in detail elsewhere. 

In summary, we have presented an experimental measurement of the
conductance distribution of mesoscopic fluctuations at a quantum Hall
transition.  The results are consistent with the hypothesis that the
distribution function is given by random matrix theory for the
single-channel unitary $S$-matrix. 

DHC is particularly grateful to Crispin Barnes for many stimulating
discussions about the possible microscopic nature of the fluctuations.  We
also thank P. Mello and M. Pepper for helpful discussions, J. T. Nicholls
for invaluable experimental assistance, and Yuki Oowaki of Toshiba for
supplying the devices.  DHC acknowledges the UK EPSRC for financial
support.


\begin{figure}
\caption{Characteristic of a $0.6 \times 0.6$~$\mu$m$^2$ MOSFET device
at $T \sim 100$~mK, at (a) $B = 0$ and (b) $B = 16$~T.  A $2.7$~k$\Omega$
series resistance was subtracted from the data in (b).  Inset: idealized
sketches of a device in cross-section and plan view.}
\end{figure}

\begin{figure}
\caption{(a) Sample of the universal conductance fluctuations at low
magnetic field.  (b) Histogram obtained from many similar datasets at
different $V_g$, normalized to a maximum value of unity.  The solid curve
is the best Gaussian fit. (c) The first quantum Hall transition from
Figure 1 (b).  The thick solid line is the same data smoothed over $16$~mV
in $V_g$.  (d) Histogram of the region of datapoints joined with a thin
solid line in (c), normalized to an average value of unity.}
\end{figure}

\begin{figure}
\caption{Further examples of quantum Hall transitions at $B = 16$ T.  Each
has been offset arbitrarily in $V_g$, for convenient comparison, but not 
in $G$.  The device dimensions, in the form length~$\times$~width, are 
indicated in microns.}
\end{figure}


\begin{thebibliography}{99}
\bibitem{sup} At magnetic fields below about 10 mT the aluminium contacts
went superconducting, making determination of the phase coherence length
using weak localization studies impracticable. Measurements nominally at
$B=0$ were actually done at $B=20 mT$.
\bibitem{bev} The evolution with $B$ is so slow that it is not possible to
significantly decrease the error in the measured distribution by combining
datasets taken at different magnetic fields.
\end{thebibliography}
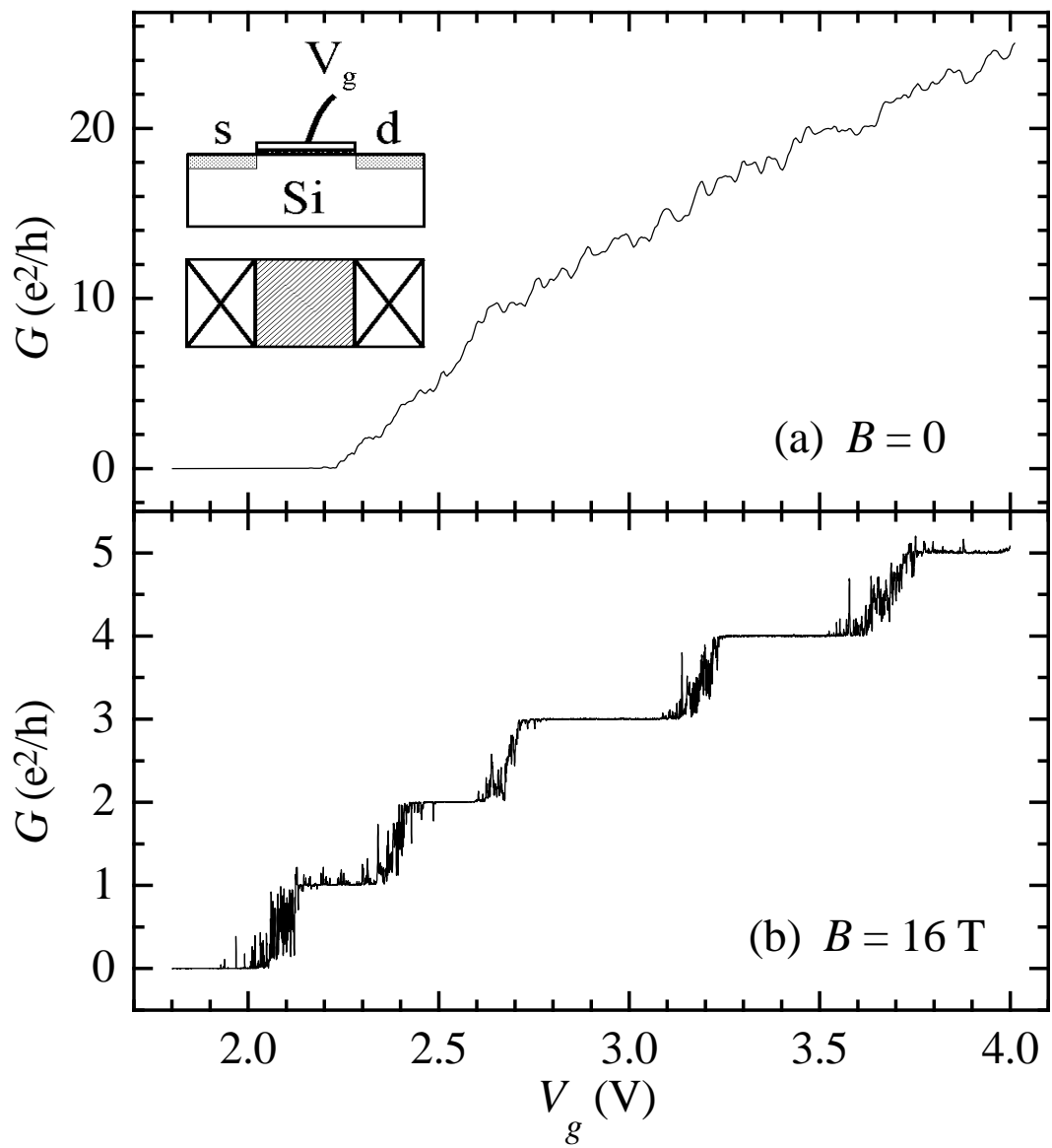
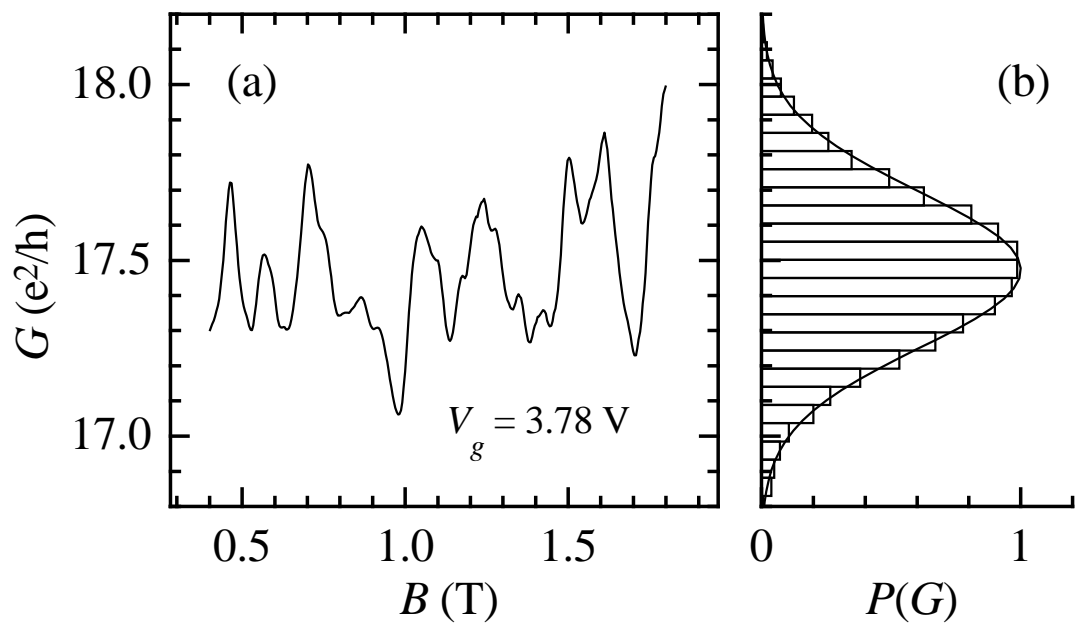
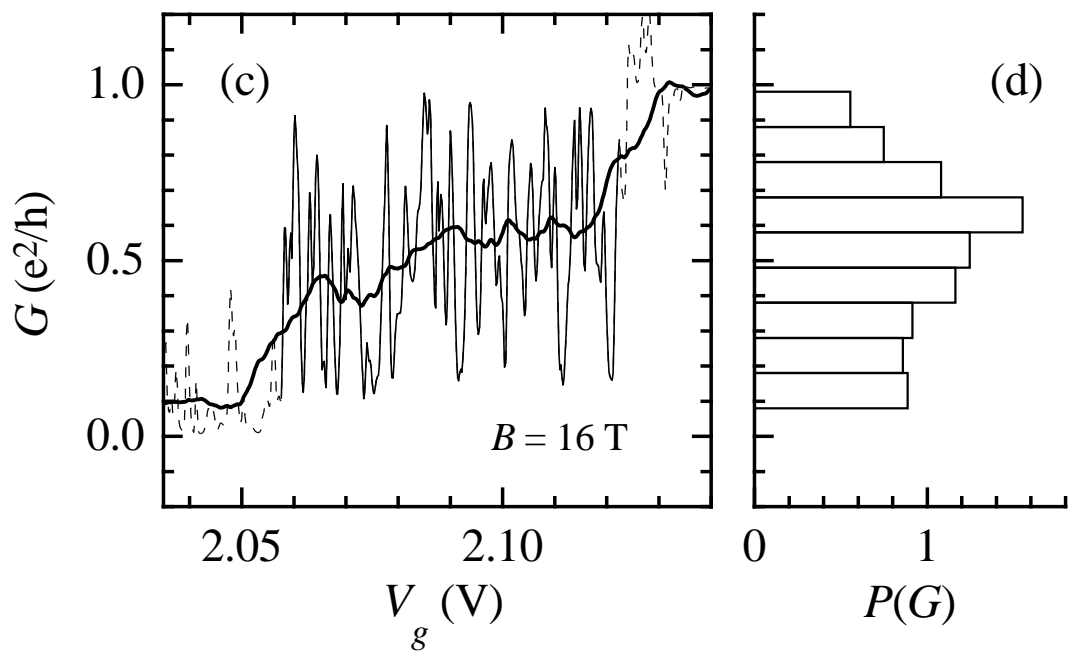
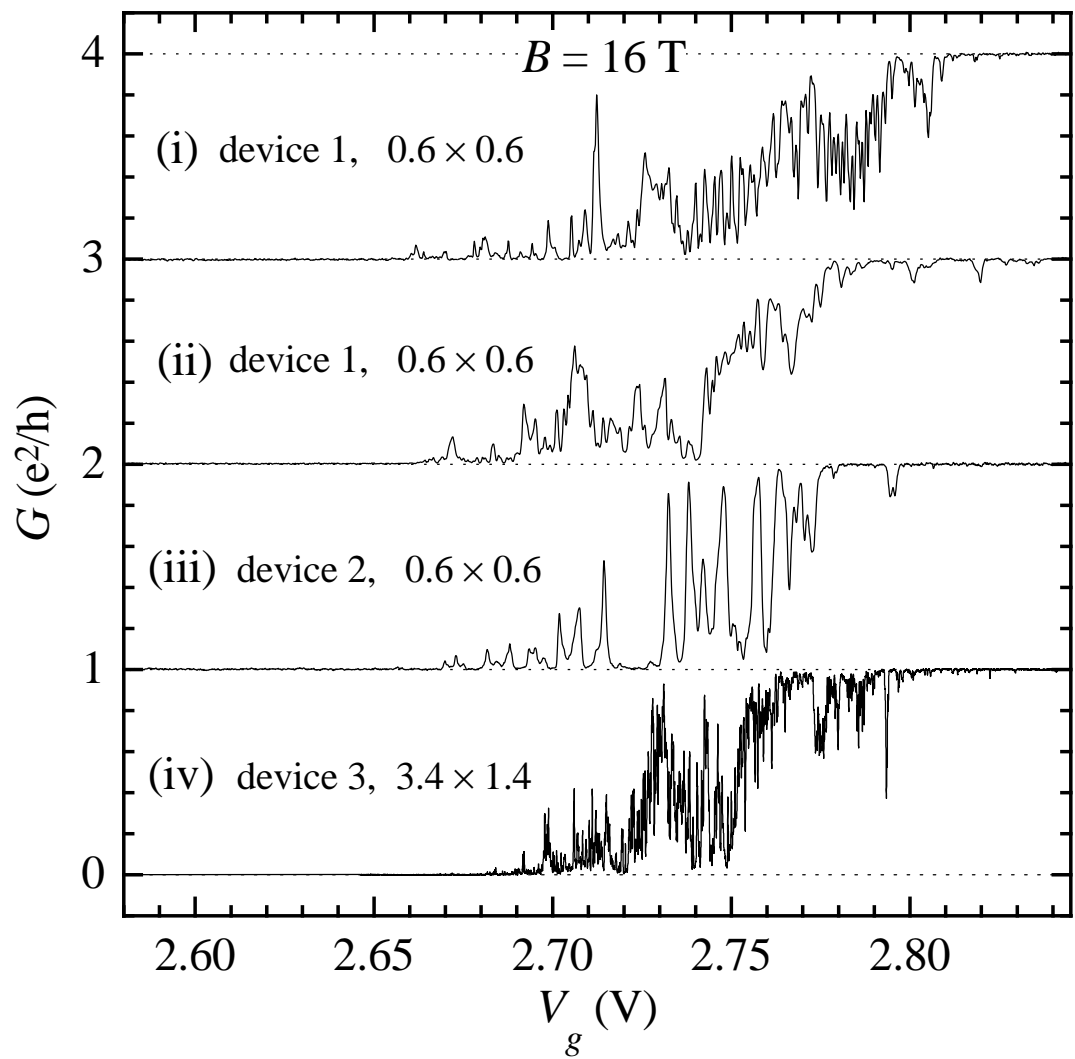
\end{document}